\documentclass[twocolumn,showpacs,preprintnumbers,amsmath,amssymb]{revtex4}
\usepackage{amsmath}
\usepackage{graphicx}% Include figure files
\usepackage{dcolumn}% Align table columns on decimal point
\usepackage{bm}% bold math

\begin{document}
%\preprint{CREAM/Manuscript-Sub-2007-GR-01}

\title{Atom--photon momentum entanglement with quantum interference}% Force line breaks with \\

\author{Rui Guo}
\author{Hong Guo}\thanks{Author to whom correspondence should be
addressed. E-mail: hongguo@pku.edu.cn, phone: +86-10-6275-7035.}

\affiliation{CREAM Group, School of Electronics Engineering and
Computer Science, Peking University, Beijing
100871, China\\}%

\date{\today}% It is always \today, today,

\begin{abstract}
With quantum interference of two-path spontaneous emissions, we
propose a novel scheme to coherently control the atom--photon
momentum entanglement through atomic internal coherence. A novel
phenomenon called ``momentum phase entanglement'' is reported, and
we found, under certain conditions, that more controllable entangled
state can be produced with super--high degree of entanglement.
\end{abstract}

.

%\keywords{Use showkeys class option if keyword display desired}%
\maketitle

\emph{Introduction.}--- Entanglement with continuous variable has
fundamental importance in quantum nonlocality \cite{EPR} and in
quantum information \cite{rmp}. Being one of the ways of physical
realizations, momentum entanglement plays a unique role in recent
studies \cite{Singlephoton,3-D
spontaneous,scattering,GR,exp,GR-disentanglement,photoionization,phase
entang.}. In spontaneous emission process, well localized
atom--photon entangled wavepacket \cite{Singlephoton} can be
produced due to the momentum conservation, with the degree of
entanglement inversely proportional to the linewidth of the
transition \cite{Singlephoton,3-D spontaneous,scattering,GR}.
Therefore, it is believed, by squeezing the effective transition
linewidth, that highly entangled EPR--like state \cite{EPR} could be
produced in free space \cite{scattering,GR}.

Insofar experiments \cite{exp}, the entanglement information can be
extracted by correlated momentum measurements, and it is found that
the degree of entanglement is completely detectable with the
conditional--unconditional variance ratio [$R$--ratio in Eq.
(\ref{R-def})] for a large variety of physical processes
\cite{Singlephoton,3-D
spontaneous,scattering,GR,exp,GR-disentanglement}. Therefore, it is
straightforward to ask if it is physically possible to produce
entanglement beyond this momentum detection, the present work will
answer this question. In a typical nearly--degenerated three--level
atom, which will be discussed in the following, we find that the
interference between different quantum pathways \cite{SGC,atomic
coherence contr} produces significantly the so--called
``phase--entangled'' state, in which the entanglement information
can not be evaluated properly by solely using the momentum detection
measure, i.e., the $R$--ratio. Within this model, we find that not
only the degree but also the modes of the entanglement can be
effectively manipulated by controlling the atomic internal
coherence, and the entanglement degree exhibits, ``anomalously'', to
be proportional to the atomic linewidth of the excited energy
levels. Therefore, it is possible to use this proposed scheme to
produce a novel and more controllable highly entangled atom--photon
system in realistic applications.
\begin{figure}
\centering
\includegraphics[height=2.8cm]{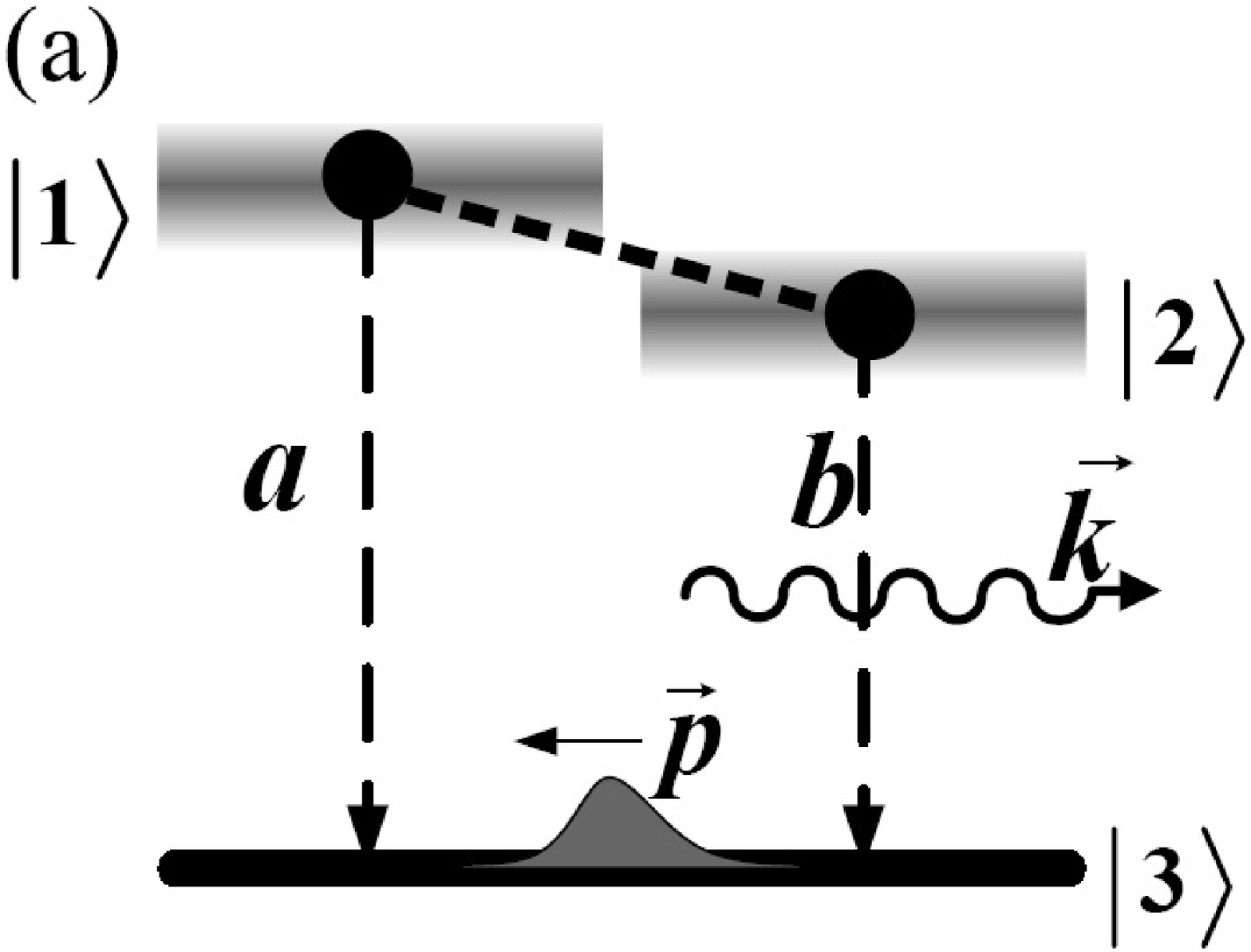}\ \ \ \ \ \
\includegraphics[height=2.8cm]{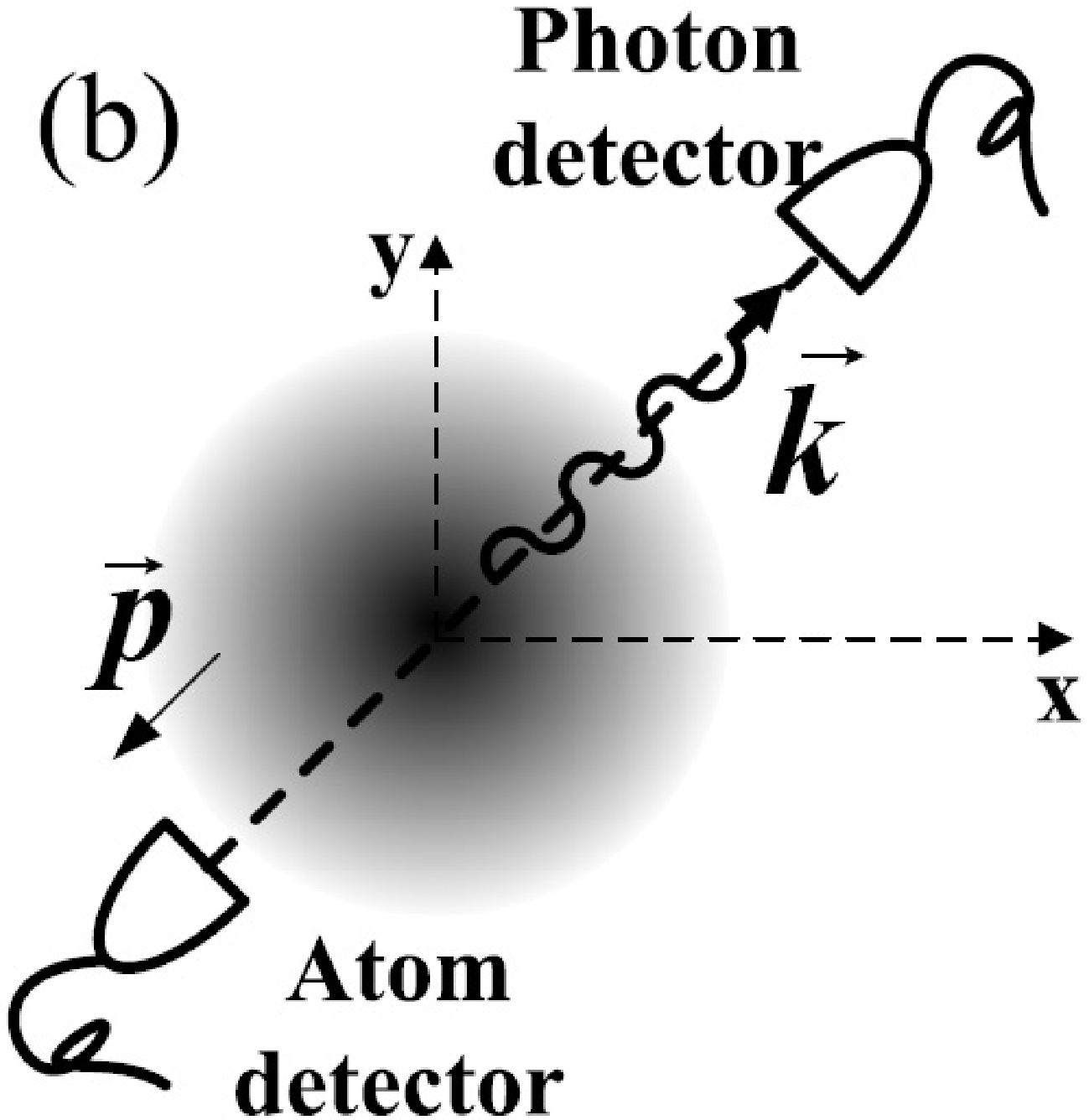}
\caption{(a) The atom has two closely--lying upper levels and
provides two quantum pathways for the spontaneous emissions. The
momentum conservation leads to the entanglement of the emitted
photon and the recoiled atom. (b) Schematic diagram for the momentum
detection. The two detectors are fixed in one dimension as in the
reported experiments \cite{exp}.}
\end{figure}

\emph{Theoretical model.}--- As shown in Fig. 1 (a), the
three--level atom has two transition pathways ``a'' and ``b'' to
induce the momentum entanglement with the emitted photon due to
momentum conservation. We will, in the following, only consider the
strong interference conditions, which assumes that the dipoles
$\vec{\mu}_{a,b}$ parallel with each other \cite{SGC}, i.e.,
$\varepsilon\equiv
\vec{\mu}_{a}\cdot\vec{\mu}_{b}/|\vec{\mu}_{a}|\cdot|\vec{\mu}_{b}|=1$,
and the upper-levels are nearly-degenerated: $\omega_{12}\equiv
\omega_{a}-\omega_{b}<\gamma_{a,b}\ll \omega_{a,b}$, where
$\gamma_{a,b}$ and $\omega_{a,b}$ are the linewidths and central
frequencies of the two transitions, respectively. The Hamiltonian of
this system with the rotating wave approximation is:
\begin{eqnarray} \hat{H}&=&\frac{(\hbar\hat{\vec{q}})^{2}}{2m}+\sum_{\vec{k}}\hbar
\omega_{\vec{k}}\hat{a}^{\dag}_{\vec{k}}\hat{a}_{\vec{k}}+\hbar
\omega_{a}\hat{\sigma}_{11}+\hbar \omega_{b}\hat{\sigma}_{22}\\
\nonumber
 &+&\hbar\sum_{\vec{k}}\left[g_{a}(\vec{k})\hat{\sigma}_{31}\hat{a}^{\dag}_{\vec{k}}e^{-i\vec{k}\cdot\vec{r}}
+g_{b}(\vec{k})\hat{\sigma}_{32}\hat{a}^{\dag}_{\vec{k}}e^{-i\vec{k}\cdot\vec{r}}+{\rm
H.c.} \right],
\end{eqnarray}
where $\hbar\hat{\vec{q}}$ and $\vec{r}$ are the atomic
center--of--mass momentum and position operators, $m$ is the atomic
mass, $\hat{\sigma}_{ij}$ is the atomic operator,
$\hat{a}_{\vec{k}}$ ($\hat{a}^{\dag}_{\vec{k}}$) is the annihilation
(creation) operator for the $k$th vacuum mode with wave vector
$\vec{k}$ and frequency $\omega_{\vec{k}}\equiv ck$, and
$g_{a,b}(\vec{k})$ are the coupling coefficients for the two
transitions, where we use $\vec{k}$ to denote both the momentum and
polarization for simplicity.

It is convenient to depict the above interaction system in
Schr\"{o}dinger picture and expand the photon--atom state as:
\begin{eqnarray}
\nonumber &&|\psi\rangle = \sum_{\vec{q},n=1,2}\exp \left [-i \left
( \frac{\hbar q^{2}}{2m}+\omega_{a} \right )t \right ]
A_{n}(\vec{q},t)
|\vec{q},0,n\rangle \\
&&+ \sum_{\vec{q},\vec{k}} \exp \left [-i \left ( \frac{\hbar
q^{2}}{2m}+c k \right )t \right ]
B(\vec{q},\vec{k},t)|\vec{q},1_{\vec{k}},3\rangle\ ,
\end{eqnarray}
where the arguments in the kets denote, respectively, the wave
vector of the atom, the photon, and the atomic internal states.

Using the Born--Markov approximation, the evolution of atom--photon
state can be solved from Schr\"{o}dinger equation. Suppose the atom
is initially prepared in a superposed internal state
$A_{10}|1\rangle+A_{20}|2\rangle$ with a Gaussian wavepacket
$G(\vec{q})\propto \exp{[-(\vec{q}/\delta q)^{2}]}$, and the
detections are restricted as in Fig. 1 (b), then the
one--dimensional steady state solutions yields:
\begin{eqnarray}
&&A_{1}(q,t\rightarrow\infty)= A_{2}(q,t\rightarrow\infty)=0,\\
&& B(q,k,t\rightarrow\infty)\propto \exp[-(\Delta
q/\eta)^{2}]\times \label{wavefunc} \\
&& \left[\frac{C_{1}(2g_{b}s_{1}/\varepsilon
\sqrt{\gamma_{a}\gamma_{b}}-g_{a})}{i(\Delta q\!+\!\Delta
k)\!+\!(s_{1}/\gamma_{a}\!-\!\frac{1}{2})}\!+\!
\frac{C_{2}(2g_{b}s_{2}/\varepsilon
\sqrt{\gamma_{a}\gamma_{b}}-g_{a})}{i(\Delta q\!+\! \Delta
k)\!+\!(s_{2}/\gamma_{a}\!-\!\frac{1}{2})} \right] , \nonumber
\end{eqnarray}
where the parameters are defined as:
\begin{eqnarray}
&& s_{1,2}\equiv\frac{1}{2}(\lambda
\pm\sqrt{\lambda^{2}+\varepsilon^{2}\gamma_{a}\gamma_{b}}),\ \lambda=\frac{\gamma_{a}-\gamma_{b}+2i\omega_{12}}{2},\nonumber \\
&& C_{1,2}\equiv
\pm\frac{s_{2,1}A_{10}+\frac{1}{2}\varepsilon\sqrt{\gamma_{a}\gamma_{b}}A_{20}}{s_{2}-s_{1}},
\ \eta\equiv\frac{\delta q \hbar k_{0}}{m \gamma_{a}},
\end{eqnarray}
and the effective wave vectors are defined by:
\begin{eqnarray}
\Delta k &\equiv&\frac{k-k_{0}}{\gamma_{a}/c}\ , \ \Delta q
\equiv\frac{\hbar k_{0}}{m \gamma_{a}}(q-k_{0})\ ,\
k_{0}=\frac{\omega_{a}}{c}.
\end{eqnarray}

\begin{figure}
\centering
\includegraphics[height=3.6cm]{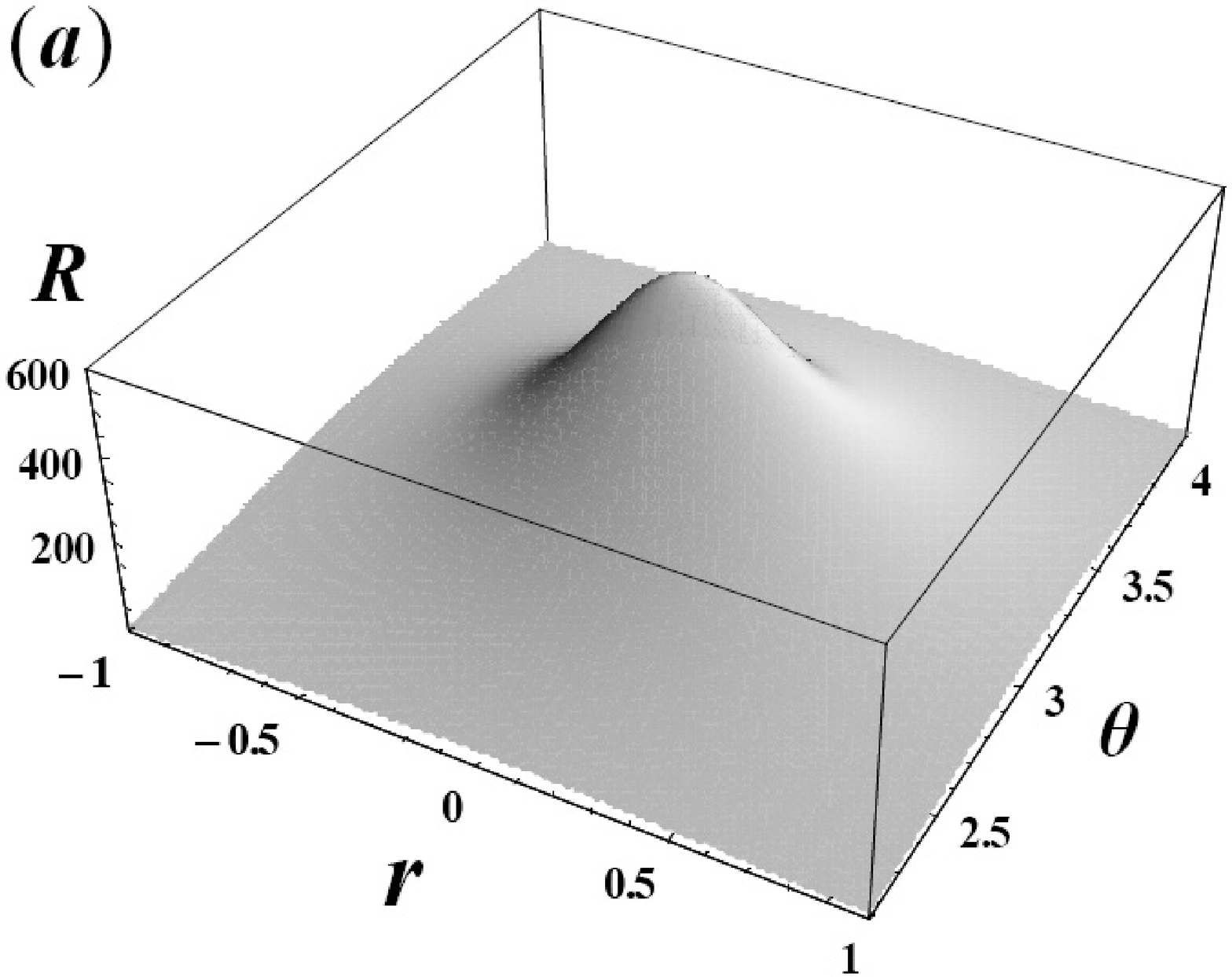}\includegraphics[height=3.6cm]{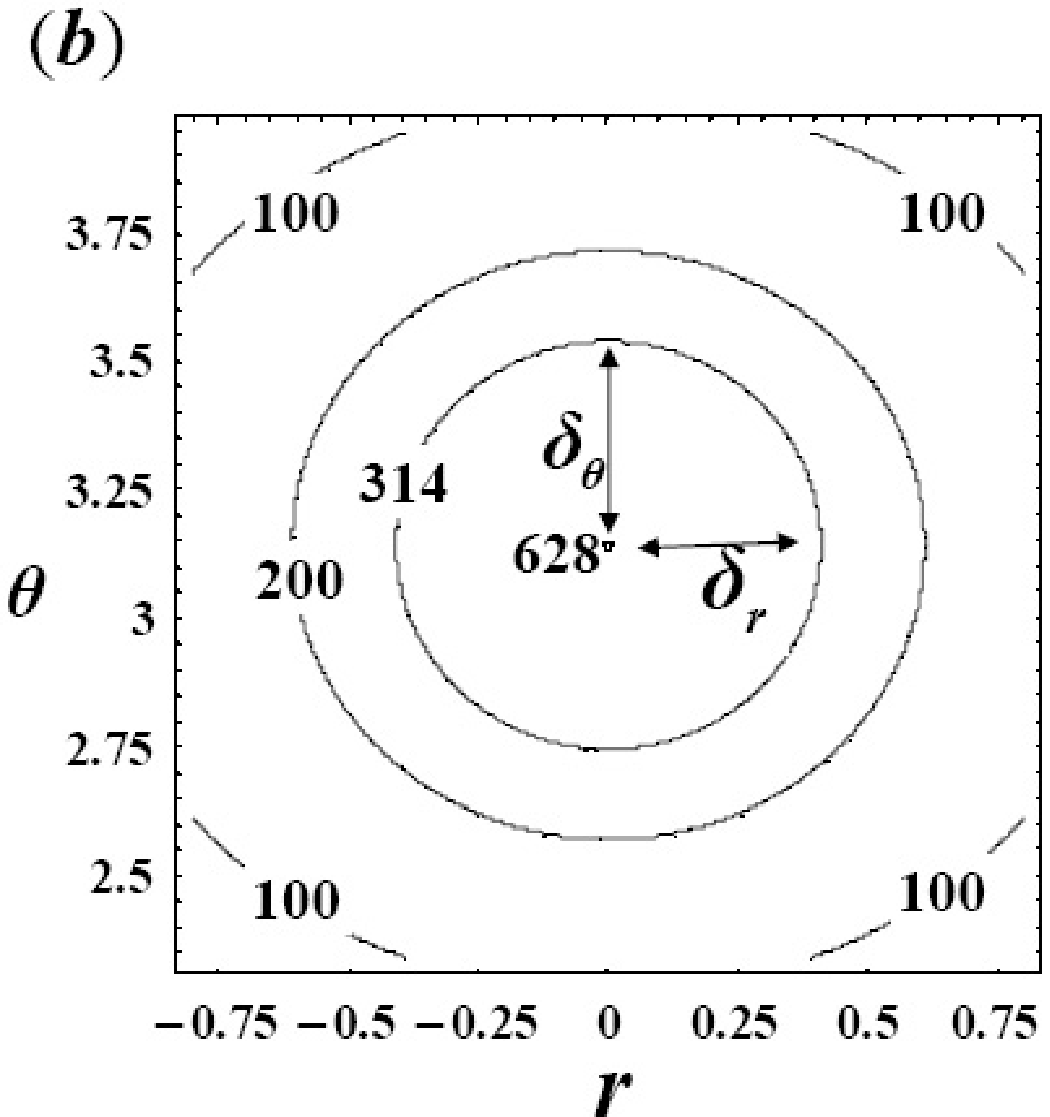}
\includegraphics[height=3.5cm]{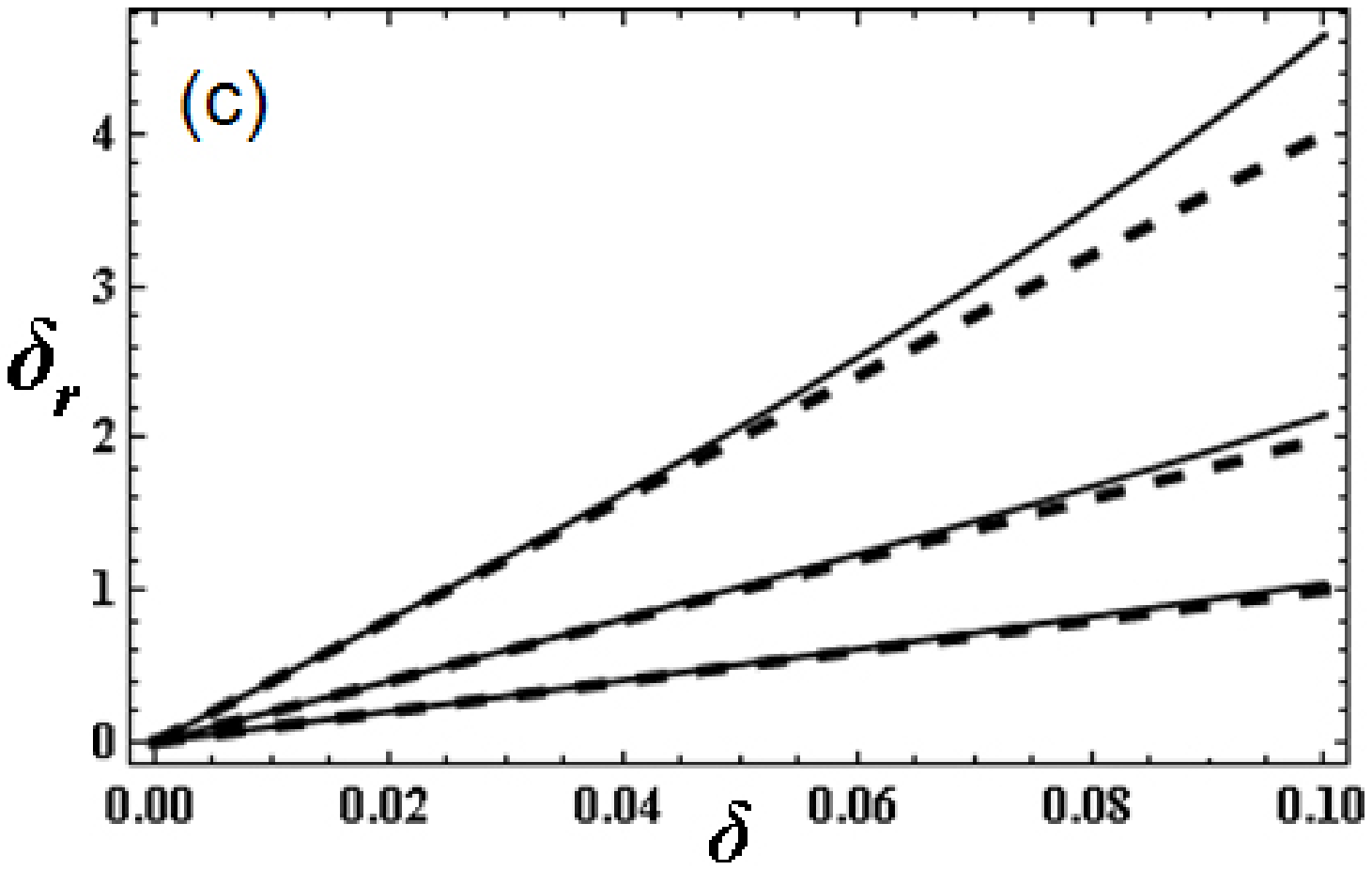}
\caption{(a) The ``amplitude entanglement'' $R$--ratio is plotted in
dependence on the atomic coherence $r$ and $\theta$ with
$\delta=0.02$, $\eta=0.1$. (b) The contour plot of Fig. 2 (a). The
circular contours indicate the symmetric roles played by $r$ and
$\theta$ in controlling the $R$--ratio. The FWHM is denoted by
$\delta_{\theta}$ and $\delta_{r}$ as in the figure. (c) The FWHM of
$R(r)$ is plotted in solid lines in dependence of $\delta$ with
$\eta=0.05, 0.1, 0.2$ from the top to the bottom. Dashed lines are
the fitted function $2\delta/\eta$.}
\end{figure}

\emph{Entanglement detection}---The nonfactorization of the
wavefunction in Eq. (\ref{wavefunc}) reveals the atom--photon
entanglement. In both theoretical \cite{3-D
spontaneous,photoionization} and experimental studies \cite{exp},
the ratio of the conditional and unconditional variances [i.e., the
$R$--ratio defined in Eq. (\ref{R-def})] plays a central role, since
it is a direct experimental measure of the nonseparability
(entanglement) of the system.

With the single--particle measurement, the unconditional variance
for the effective atomic momentum is determined as $\delta q^{{\rm
single}}= \langle \Delta q^{2} \rangle-\langle \Delta q
\rangle^{2}=\int {\rm d} \Delta q\ {\rm d} \Delta k \Delta q^{2}
|B(q,k)|^{2}$, where the average $\langle\cdot\rangle$ is taken over
the whole ensemble. Meanwhile, the coincidence measurement gives the
conditional variance as $\delta q^{{\rm coin}}= \langle \Delta q^{2}
\rangle_{\Delta k_{0}} -\langle \Delta q \rangle^{2}_{\Delta
k_{0}}\propto \int {\rm d} \Delta q\ \Delta q^{2} |B(q,\Delta
k_{0})|^{2}$, where the photon is previously detected at some known
$\Delta k_{0}$. With these two variances, the entanglement is
evaluated by:
\begin{eqnarray}
R\equiv\delta q^{{\rm single}}/\delta q^{{\rm coin}}\geq 1.
\label{R-def}
\end{eqnarray}

Due to the interference of two transition pathways, the $R$--ratio
highly depends on the initial coherence of the two upper atomic
levels, which is evaluated by $A_{10}/A_{20}\equiv
\exp{(r+i\theta)}$, where $r$ depicts the relative occupation
probabilities of the two upper levels, and $\theta$ determines their
coherence phase. In further discussions, we assume
$\gamma_{a}=\gamma_{b}\equiv \gamma$, and define $\delta\equiv
\omega_{12}/\gamma<1$ for simplicity.

Following the above analysis, the dependence of $R$ on $r$ and
$\theta$ is illustrated in Fig. 2. Under the conditions $\eta\ll 1$
and $\delta^{2}/\eta \ll 1$ \cite{explanations}, we find that
$R(r,\theta)$ can well be approximated by Lorentzian function, and
the parameters $r$ and $\theta$, in spite of their quite different
physical essences, play very symmetric roles in controlling the
detectable $R$--ratio [Fig. 2 (b)]. Under these conditions, the
$R$--ratio is maximized at the dark state coherence
[$(|1\rangle-|2\rangle)/\sqrt{2}$]:
\begin{eqnarray}
R_{{\rm max}}=R(r=0, \theta=\pi)\approx \sqrt{2\pi}\eta/\delta^{2},
\label{5}
\end{eqnarray}
the full width at half maximum (FWHM) of which can be well
approximated by $\delta_{r}\approx\delta_{\theta}\approx
2\delta/\eta$, as shown in Figs. 2 (b) and (c). Therefore, with
properly chosen atomic parameters $\eta$ and $\delta$, this scheme
could be used to produce significant detectable entanglement in a
relatively large range of the initial atomic coherence. For example,
with $\eta=\delta=0.01$, the system with $R>100$ can be produced
within the range of $0.018<|A_{10}/A_{20}|^{2}=e^{2r}<55$, and
$0.36\pi <\theta<1.6\pi$.
\begin{figure}
\centering
\includegraphics[height=3.3cm]{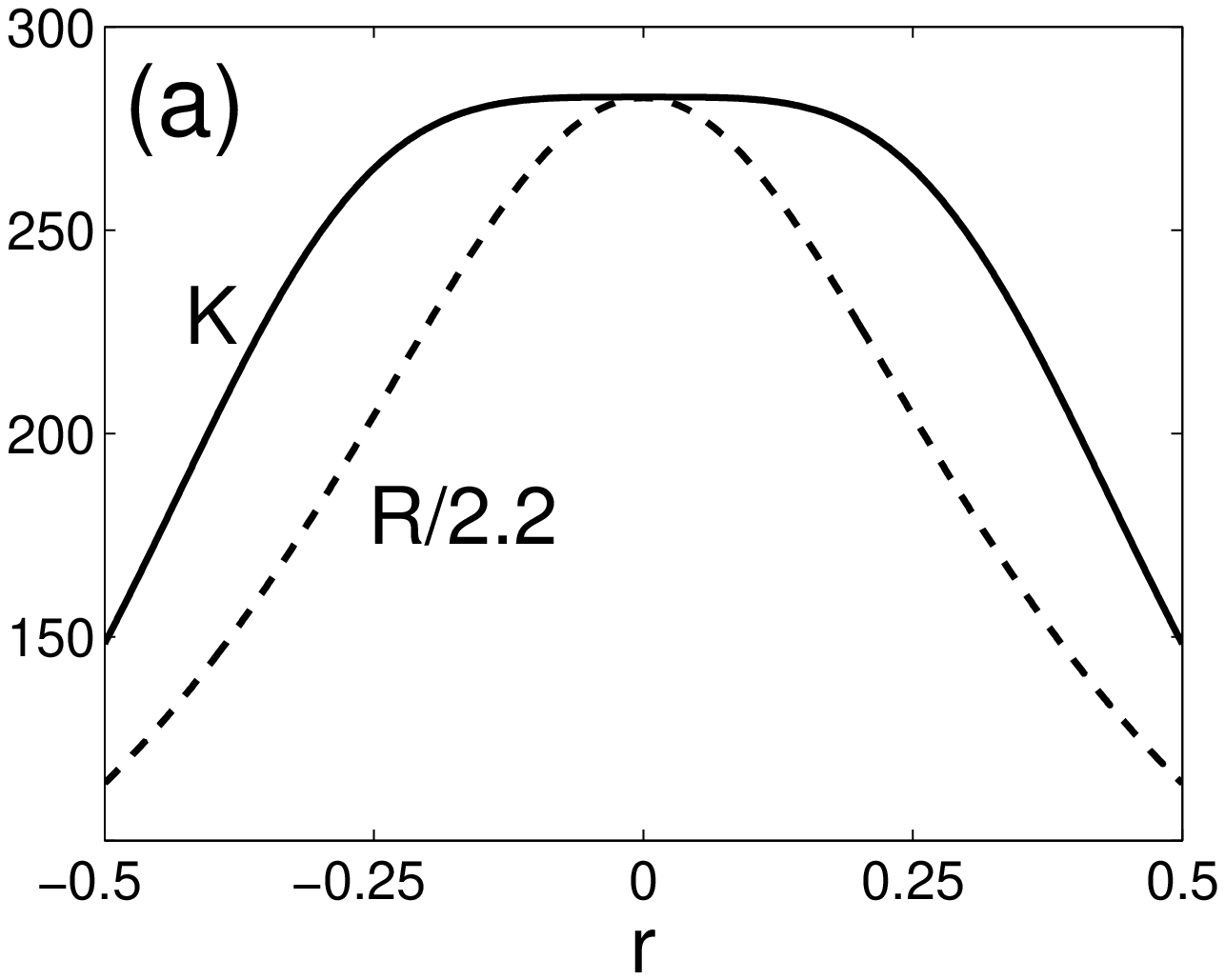}\ \ \ \includegraphics[height=3.3cm]{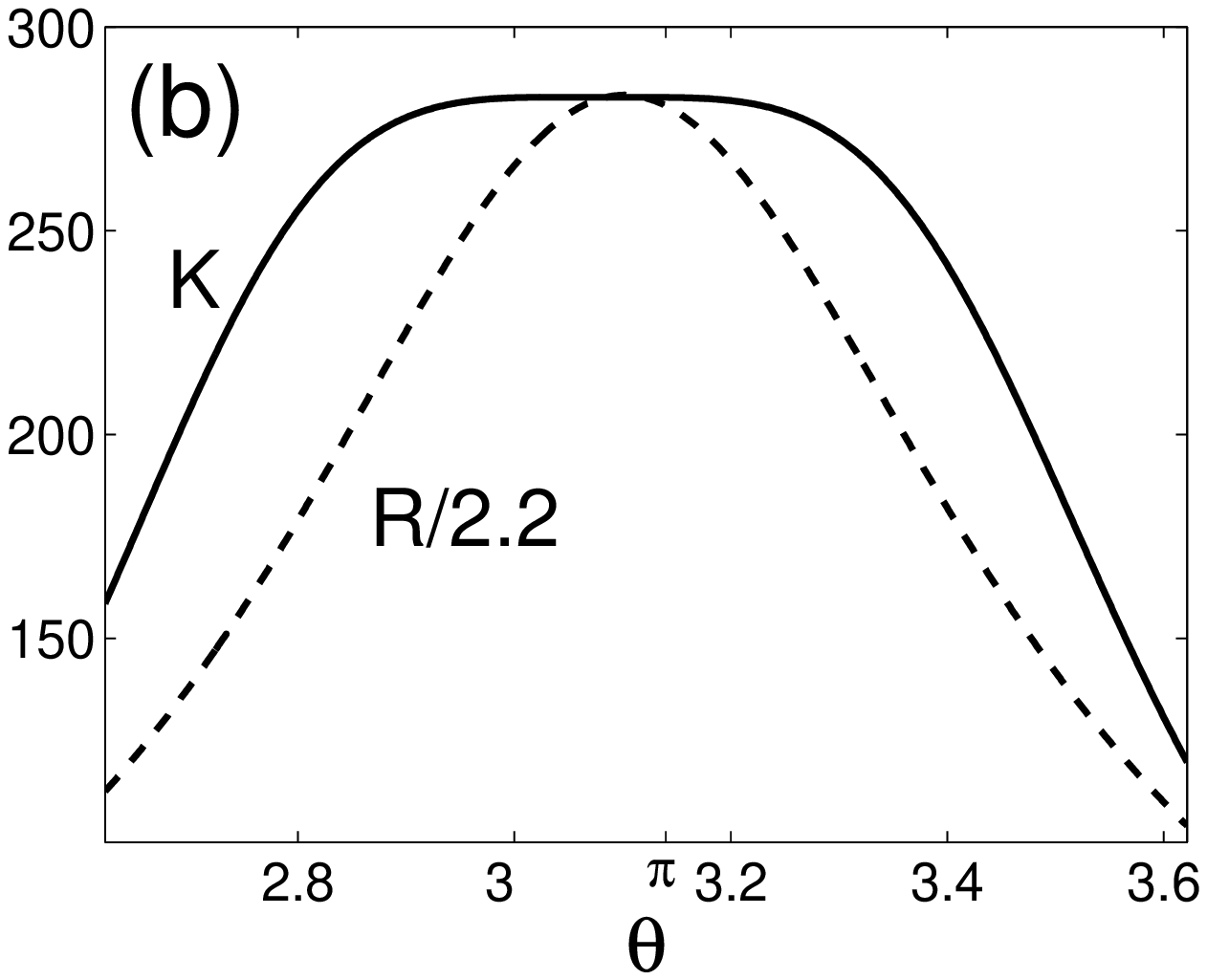}
\includegraphics[height=3.7cm]{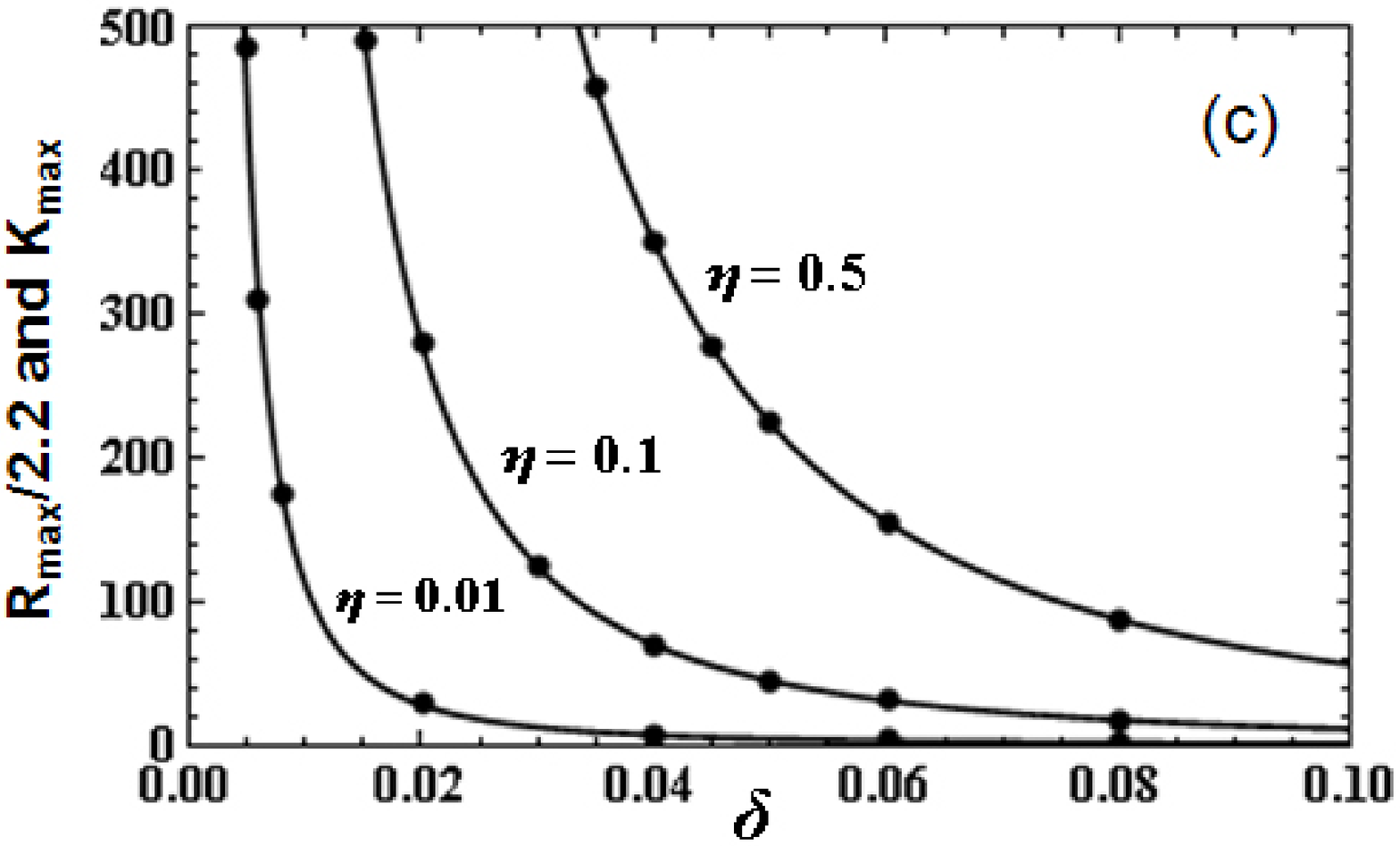}
\caption{Plots of $K$ and $R/2.2$ in dependence on the atomic
coherence $r$ or $\theta$, where $\eta=0.1$, $\delta=0.02$. (a)
$\theta$ is fixed at $\pi$. (b) $r$ is fixed at $0$. (c) $K$ (in
spots) and the function $R/2.2$ (in solid line) are plotted at the
dark--state coherence $r=0$ and $\theta=\pi$.}
\end{figure}

\emph{Phase entanglement.}--- For a bipartite pure--state system,
the degree of entanglement can be completely evaluated by the
Schmidt number \cite{Singlephoton,Parametric Down Conversion}
\begin{eqnarray}
K\equiv 1/\sum_{n}\lambda_{n}^{2}\geq 1 \label{Schmidt num},
\end{eqnarray}
where $\lambda_{n}$'s are eigenvalues for the entangled atomic modes
$\psi_{n}(q)$ and photonic modes $\phi_{n}(k)$ in the Schmidt
decomposition \cite{Schmidt num}:
$B(q,k)=\sum_{n}\sqrt{\lambda_{n}}\psi_{n}(q)\phi_{n}(k)$.

In previous studies on atom--photon momentum entanglement
\cite{Singlephoton,3-D spontaneous,scattering,GR}, the $R$--ratio
well measures the entanglement since one has $R\propto K$. However,
this is not true when the quantum interference is strong as shown in
this model: since the $R$--ratio is constructed from the module part
of the wavefunction, it reveals only the amplitude correlation
between two particles' momentum; therefore, when the phase is
critical for the nature of the entanglement due to the interference,
the traditional $R$--ratio measurement becomes inadequate for
detecting the full entanglement information. Actually, by
controlling the interference with the atomic internal coherence, one
may produce two systems with $K>K'$ whereas $R<R'$, which indicates
that significant entanglement information may be lost by the
momentum detection with only the $R$--ratio.

We compare $K$ and $R$ in Fig. 3, from which one finds that both of
them are maximized at the dark-state coherence $(r=0,\ \theta=\pi)$.
However, compared with $R(r,\theta)$, $K(r,\theta)$ exhibits a much
slower decay in the vicinity of the maximum, which indicates that,
with different initial atomic coherences, some entanglement
information may be transferred into the ``phase'' and can not be
measured only by the amplitude--based detections. This phenomenon is
particularly important for some highly entangled states, e.g., under
the condition $\delta=2\times 10^{-3},\ \eta=0.1,\ r=0,\
\theta=\pi$, one may prepare a highly entangled state with
$K\approx2.8\times 10^{4}$ and $R\approx 6.2\times 10^{4}$; however,
when the initial atomic momentum and coherence change to
$\eta'=2\eta, r'=0.13$, the entanglement of the system does not
change, i.e., $K'= K$, but the $R$-ratio detection shows that $R' =
0.05 R$. This shows that some entanglement information of the system
is transferred into the phase.

Similar phenomenon of the so--called ``phase entanglement'' has been
reported recently in the position space \cite{3-D
spontaneous,photoionization,phase entang.}: due to the spreading of
the wavepacket, it appears instantly and must be detected by a
series of spatial measurements in time \cite{photoionization}. For
the momentum ``phase entanglement'' in this scheme, however, since
it is caused by the quantum interference and is not affected by the
wavepacket's spreading, this phenomenon keeps steady in time and
could be much easier to be directly observed in experiments
\cite{exp}.

It is possible to evaluate the ``phase entanglement'' for the highly
entangled states in this scheme. For the entanglement maximized at
the dark--state coherence, the wavefunction takes a similar form as
if no interference occurs \cite{Singlephoton,3-D
spontaneous,scattering,GR}:
\begin{eqnarray}
B(q,k,t\rightarrow\infty)\propto \frac{\exp{[-(\Delta q/\eta)^{2}}]}
{i(\Delta q+\Delta k )-\delta^{2}/4}\ , \label{7}
\end{eqnarray}
and then the Schmidt number yields:
\begin{eqnarray}
K_{{\rm max}}\approx 1+0.28(4\eta/\delta^{2}-1). \label{4}
\end{eqnarray}
Together with Eq. (\ref{5}), one yields the relation
\begin{eqnarray}
K_{{\rm max}}=K(r=0,\theta=\pi)\approx\frac{R_{{\rm max}}}{2.2}
\approx \frac{1.12\hbar k_{0}\delta q\gamma}{m \omega_{12}^{2}},
\label{K-and-R}
\end{eqnarray}
well fulfilled for $\eta/\delta^{2}\gg1$ and $\eta\ll 1$
\cite{explanations}, as shown in Fig. 3 (c). The linear relation
between $K_{{\rm max}}$ and $R_{{\rm max}}$ shown in Eq.
(\ref{K-and-R}) indicates that the entanglement is completely
detectable with fixed dark--state coherence $r=0$ and $\theta=\pi$.
For general conditions (where $r$ and $\theta$ take arbitrary
values, see Fig. 3), we have $K\geq R/2.2$. Therefore, the degree of
``phase entanglement'' can be evaluated by the following parameter:
\begin{eqnarray}
PE\equiv 2.2K/R \geq 1,
\end{eqnarray}
which is valid for the states produced with different control
parameters $\eta$, $\delta$, $r$ and $\theta$.

A traditional idea to enhance the entanglement of momentum is to
correlate the atom and photon momentum by squeezing the transition
linewidth since $K\propto 1/\gamma$ \cite{Singlephoton,3-D
spontaneous,scattering,GR}. However, in our proposed scheme, which
employs an essentially different mechanism for producing the
entanglement through quantum interference, we have, anomalously,
that $K_{{\rm max}}\propto \gamma$ as shown in Eq. (\ref{K-and-R}).
With broader linewidth of the two upper energy levels, the
interference will be enhanced and, as a result, increases the
momentum entanglement. Therefore, it is possible to use this
mechanism to produce super--high momentum entanglement even for the
atomic system with large damping rate $\gamma$.
\begin{figure}
\centering
\includegraphics[height=6.8cm]{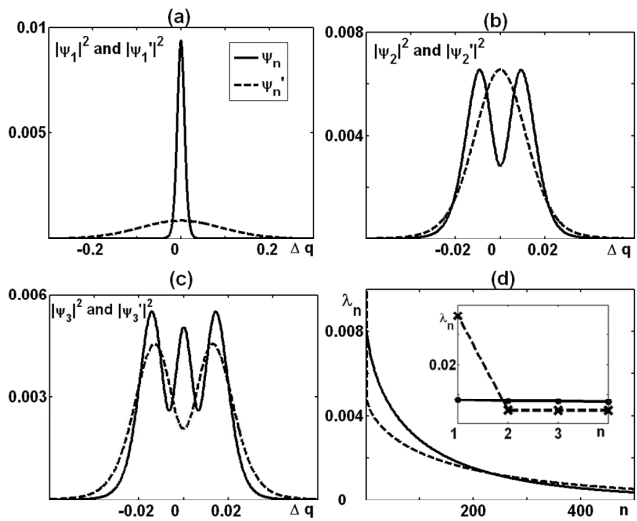}
\caption{(a) to (c) First three Schmidt modes are compared between
states $B(q,k)$ with $\delta=0.02$, $\eta=0.12$, $r=0$, $\theta=\pi
$ and $B'(q,k)$ with $\delta'=\delta$, $\theta'=\theta$,
$\eta'=0.2$, $r'=0.38$, where one has $K'= K$ and $R'\approx 0.38R$.
The phase entanglement of $B'(q,k)$ broadens its first Schmidt mode
and decreases the number of peaks for the rest modes. (d) The
distributions of the eigenvalues in their Schmidt decompositions,
where the inset shows the first four eigenvalues. The solid and
dashed lines are for $B(q,k)$ and $B'(q,k)$, respectively. }
\end{figure}

\emph{Entangled modes.}---The phase entanglement can, more
intuitively, be understood in terms of Schmidt decomposition.
According to Eq. (\ref{Schmidt num}), $K$ is a measure for the
number of the important Schmidt modes, while $R$--ratio is related
to the coherence between these modes [$\psi_{n}(q)$]', which can be
seen more clearly by rewriting the unconditional and conditional
variances as: $ \delta q^{{\rm single}}= \int {\rm d} \Delta q\
\Delta q^{2}\sum_{n}\lambda_{n}|\psi_{n}(q)|^{2} = \langle\Delta
q^{2}\rangle_{E_{{\rm i}}}$, $\delta q^{{\rm coin}}\approx \int {\rm
d} \Delta q\ \Delta q^{2}
|\sum_{n}\sqrt{\lambda_{n}}\psi_{n}(q)|^{2}=\langle\Delta
q^{2}\rangle_{E_{{\rm c}}}$. These formulae indicate that the
unconditional variance $\delta q^{{\rm single}}$ is the variance
taken by an ``incoherent'' superposition of different Schmidt modes
weighed by $\lambda_{n}$, i.e., $E_{{\rm i}}(q)\equiv
\sum_{n}\lambda_{n}|\psi_{n}(q)|^{2}$, while the conditional
variance $\delta q^{{\rm coin}}$ is taken over a ``coherent
superposition'' of different modes, i.e., $E_{{\rm c}}(q)\equiv
|\sum_{n}\sqrt{\lambda_{n}}\psi_{n}(q)|^{2}$. Therefore, the
$R$--ratio defined in Eq. (\ref{R-def}) actually represents the
wavepacket narrowing caused by the coherence between different
Schmidt modes.

In Fig. 4, we compare the atomic Schmidt modes between the states
$B(q,k)$ and $B'(q,k)$ with $K'= K$ and $R'\approx 0.38 R$. It can
be seen that the phase entanglement significantly broadens the first
few Schmidt modes and decreases the number of peaks for the rest
ones; moreover, the coherence between different Schmidt modes
diminishes and, as a result, decreases the $R$--ratio. The photonic
Schmidt modes exhibit similar properties as atomic modes and remain
the Gaussian localization properties \cite{Singlephoton,scattering}
in spite of the shape distortions caused by the interference.
Therefore, it is possible to apply this scheme to efficiently
control the entangled modes under a certain degree of entanglement.

\emph{Conclusion.}--- In summary, we investigate the atom--photon
momentum entanglement caused by the quantum interference in a
three--level atom. The novel phenomenon of ``momentum phase
entanglement'' is shown and evaluated quantitatively. Using this
scheme, a novel atom-photon entangled state can be produced with
super--high degree of entanglement and controllable entangled modes.
Since the proposed configuration has been extensively studied both
theoretically and experimentally \cite{SGC,atomic coherence contr},
and can be realized by mixing different parity levels or by using
dressed-state ideas, these new features are most probable to be
examined in experiments and used in realistic applications
\cite{rmp}.

This work is supported by the National Natural Science Foundation of
China (Grant No. 10474004), National Key Basic Research Program
(Grant No. 2006CB921401) and DAAD exchange program: D/05/06972
Projektbezogener Personenaustausch mit China (Germany/China Joint
Research Program).


\begin{references}
\bibitem{explanations} In typical atomic system, $\eta$ is of the order $0.1$ or smaller, e.g., for sodium, one has
$\eta\approx 0.05$; moreover, the condition $\delta^{2}/\eta\ll 1$
is equivalent to $K_{{\rm max}}\gg 1$, which is our major concern.
\end{references}
\end{document}